# Measuring Lateral Magnetic Structure in Thin Films Using Polarized Neutron Reflectometry

W.-T. Lee [1], S. G. E. te Velthuis [2], G. P. Felcher [2], F. Klose [1], T. Gredig [3], D. Dahlberg [3], B. V. Toperverg [4]

[1] *Oak Ridge National Laboratory, Oak Ridge, Tennessee 37831, USA*

[2] *Argonne National Laboratory, Argonne, Illinois 60439, USA*

[3] *University of Minnesota, Minneapolis, Minnesota 55455, USA*

[4] *Forschungszentrum Jülich, D - 52425 Jülich, Germany*

email: hlee@anl.gov

## Abstract

Polarized neutron reflectometry (PNR) has long been applied to measure the magnetic depth profile of thin films. In recent years, interest has increased in observing lateral magnetic structures in a film. While magnetic arrays patterned by lithography and submicron-sized magnetic domains in thin films often give rise to off-specular reflections, micron-sized ferromagnetic domains on a thin film produce few off-specular reflections and the domain distribution information is contained within the specular reflection. In this paper, we will first present some preliminary results of off-specular reflectivity from arrays of micron-sized permalloy rectangular bars. We will then use specular reflections to study the domain dispersion of an exchange-biased Co/CoO bilayer at different locations of the hysteresis loop.

One of the techniques to study surfaces and thin films is polarized neutron reflectometry (PNR) [1-3]. PNR has conventionally been applied to study systems in which the magnetic structure consists of a stack of laterally uniform magnetic layers. The measurements reveal the depth-dependence of the magnetization, in magnitude as well as in direction. Recently, attention has also been on measuring systems with lateral structures such as magnetic domains across a surface.

In a PNR experiment, a neutron beam with wavelength $\lambda$ incident on a flat surface at a grazing angle $\theta$ (typically less than a few degrees) and neutrons may be reflected from the surface at an angle $\phi+\theta$. Here the scattering plane is perpendicular to the surface and the reflected intensity is measured as a function of the scattering angle $2\theta+\phi$. The incident beam's wave vector is $k_i = 2\pi\sin\theta/\lambda\ z + 2\pi\cos\theta/\lambda\ x$ and the reflected neutron's wave vector is $k_f = -2\pi\sin(\phi+\theta)/\lambda\ z + 2\pi\cos(\phi+\theta)/\lambda\ x$. where $z$ is perpendicular to the sample surface and $x$ is on the sample surface and the scattering plane. The reflected intensity is measured as a function of the momentum transfer, $q = k_i - k_f$.

Laterally uniform layers only gives rise to specular reflection with $\phi = 0$, i.e., $q = q_z\ z$ with $q_z = 4\pi\sin\theta/\lambda$. As $q_z$ is a variable conjugate of the depth z from the surface of the film, a scan over suitable range of $q_z$ provides information on the chemical and magnetic depth profile of the film. When the incident neutrons are polarized along a applied magnetic field $H$, and the polarization after reflection is analyzed along the same axis, four reflectivities: $R^{++}$, $R^{--}$, $R^{+-}$, $R^{-+}$ are recorded. The first (second) sign refers to the incident (reflected) neutron polarization with respect to $H$. For a ferromagnetic film with the magnetization in the plane and $H$ is in the sample surface, $R^{++} - R^{--}$ is proportional to the magnetization component along $H$; and the spin-flip intensities, $R^{+-} = R^{-+}$, come from the magnetization component perpendicular to $H$.

For systems with chemical and/or magnetic structure that are not uniform across the surface, if the length scale of the variation is smaller or comparable to the coherence length of the neutron beam (typically in the order of microns to tens of microns), significant off-specular reflections may occur. The off-specular reflections have non-zero $q_x$ component with $q_x \approx 2\pi\sin\theta\sin\phi/\lambda$. In this case, both specular and off-specular components must be analyzed to obtain the chemical/magnetic profile of the sample [4].

Fig. 1 shows a gray-scale map of the scattered intensity as a function of the scattering angle and wavelength for (a) spin-up, and (b) spin-down incident beam. The sample is an array of rectangular permalloy ($Ni_{80} Fe_{20}$) elements patterned on a silicon substrate. The elements are 100 Å thick, 2x10 µm in dimensions and are separated by 2 µm along either sides of an element. The coercive and saturation field along the long element axis (easy axis) are 12 Oe and 30 Oe, respectively. The measurements were done with H=50 Oe applied along the easy axis perpendicular to the scattering plane. There was no polarization analyzer of the reflected neutrons. The incident angle was 0.55º and the specular reflection can be seen at 1.1º. Two off-specular fringes were evident at angles above and below the specular reflection. The momentum transfer at the off-specular fringes is related to the 4 µm period of the array by $q_x = 2\pi/4$ µm. Diffraction patterns can also be seen below the 0.55º horizon. The difference in the spin-up and spin-down intensities is a clear indication of the magnetic dependence of the scattering density. Comparison with theoretical calculations of the intensity map is currently underway.

If the lateral length scale of the surface structure is large compare to the coherence length of the neutron beam, there are little off-specular intensities and the specular reflections may contain information of the lateral structure [5]. In the case of an unsaturated ferromagnetic film with in-plane magnetization and a dispersion of magnetic domains, there are simple and transparent relations between the reflectivities:

$$\frac{R^{++} - R^{--}}{R_s^{++} - R_s^{--}} = \frac{R^+ - R^-}{R_s^+ - R_s^-} = <\cos \varphi>, \qquad (1)$$

$$\frac{R^{-+}}{R_s^{-+}(90°)} = <\sin^2 \varphi>, \qquad (2)$$

where $\varphi$ is a domain's magnetization angle measured from the applied field direction, the subscript s denotes values obtained at saturation, and $R_s^{-+}(90°)$ is obtained when the magnetization is saturated perpendicular to **H**. Eq. (1) gives the normalized averaged magnetization component along **H**, which can also be measured by conventional magnetometry. Eq. (2), however, contains additional information: the dispersion of the magnetic domains. The domain dispersion can be quantified by $\chi^2 = <\cos^2\varphi> - <\cos \varphi>^2$.

The following examples illustrate the meaning of $\chi^2$: A maximum dispersion ($\chi^2=1$) at magnetization reversal ($<\cos\varphi>=0$) means the domains are either parallel or anti-parallel to the applied field, i.e., the reversal process is domain switching. A value $0\leq\chi^2<1$ at the reversal means domain rotation occurs.

As an example, we consider PNR measurements on an exchange-bias bilayer CoO(45Å)/Co(130Å). The sample showed a training effect (Fig. 2 insert) [6]: After field-cooling at +5 kOe to 10 K, which is below the exchange-bias blocking temperature at 130 K, the first magnetization reversal was different from subsequent reversals. The data denoted by the symbols in the insert of Fig. 2 were obtained from applying Eq. (1) to the measured $R^+$ (filled symbols) and $R^-$ (open symbols) shown in Fig. 2a. Measurements were done, after field cooling, at H = +5 kOe (saturation), -1050 Oe (1$^{st}$ reversal), +350 Oe (2$^{nd}$ reversal), and +450 Oe (after 2$^{nd}$ reversal). It is worth noting that $R^+ = R^-$ at the reversals. This is evident in Fig. 2a, to within the accuracy of the measurements. It is important to note that, although off-specular scattering of the Yoneda type was observed when the sample was not saturated, its intensity had never exceeded a few percent of the reflected beam, that neglecting it does not affect the conclusions reached here. Fig. 2b shows the spin-flip reflectivities. The filled circles are $R^{-+}(90°)$ measured at remnant. For this measurement, the sample was first field cooled in 2 kOe. Then the field was replaced by a few Oe applied in the perpendicular direction to keep the neutron polarization during the measurement. The magnetization at remnant for this sample was close to its saturated value. The small field applied during the measurement did not alter the magnetism of the film. $R^{-+}$ measured with the film saturated along **H** is also shown (open circle). Ideally, there would be no spin-flip intensity. The non-zero intensity came from the finite efficiency of the polarizer/analyzer and spin-flipper. Further analysis has taken into account the efficiency of these components. $R^{-+}$ were clearly different between the two reversals, suggesting that the processes were different for the initial and subsequent reversals. The spin-flip reflectivity at +450 Oe is only slightly smaller than that at the second reversal. For a better understanding of the film's magnetic domain structure, one must look at the domain dispersion $\chi^2$ (Fig. 3): At the first reversal, $\chi^2$ is close to 1. From the discussion above, this implies that the initial reversal occurred mainly through domain switching. The second reversal gave $\chi^2 \approx 0.65$. There were substantial domain rotations during the reversal. $\chi^2$ is reduced to slightly less than 0.5 when the field is increased to +450 Oe after the second reversal. This agrees with the picture that, as the field increases, more domains are aligned towards the field direction. A comparison of these quantitative results of the domain dispersion with theoretical calculations, for instant, the Stoner-Wohlfarth model, is the subject of further investigations.

In conclusion, we have shown how PNR can be applied to study lateral magnetic structures on a surface. For structures with small length scale, we have shown the off-specular scattering from arrays of micron-sized magnetic elements. For large length scale structures, we have illustrated how to obtain the ferromagnetic domain dispersion of an exchange biasing CoO/Co bilayer film along the hysteresis loop.

Work done at Argonne National Laboratory was supported by US DOE, Office of Science contract #W-31-109-ENG-38 and by Oak Ridge National Laboratory, managed for the U.S. D.O.E. by UT-Battelle, LLC under Contract No. DE-AC05-00OR22725. Work done at the University of Minnesota was supported by the NSF MRSEC NSF/DMR – 9809364.

**FIGURE CAPTIONS**

Fig. 1  Scattered intensity map for (a) spin-up, and (b) spin-down incident beam from a sample consisted of arrays of micron-sized magnetic elements on a silicon surface.

Fig. 2  (a) $R^+$ (filled symbols) and $R^-$ (open smbols); (b) $R^{-+}$ of a CoO/Co bilayer at 10 K. At H = -1050 Oe (upper triangle), +350 Oe (lower triangle), +450 Oe (square). Also, in (a), at H=+5 kOe (circle); and in (b), $R^{-+}$ (90°) (filled circle) and $R_s^{-+}$ (open circle).

Fig. 3  Magnetic domain dispersion $\chi^2$ of the CoO/Co bilayer at 10 K and H = -1050 Oe (upper triangle), +350 Oe (lower triangle), and +450 Oe (square).

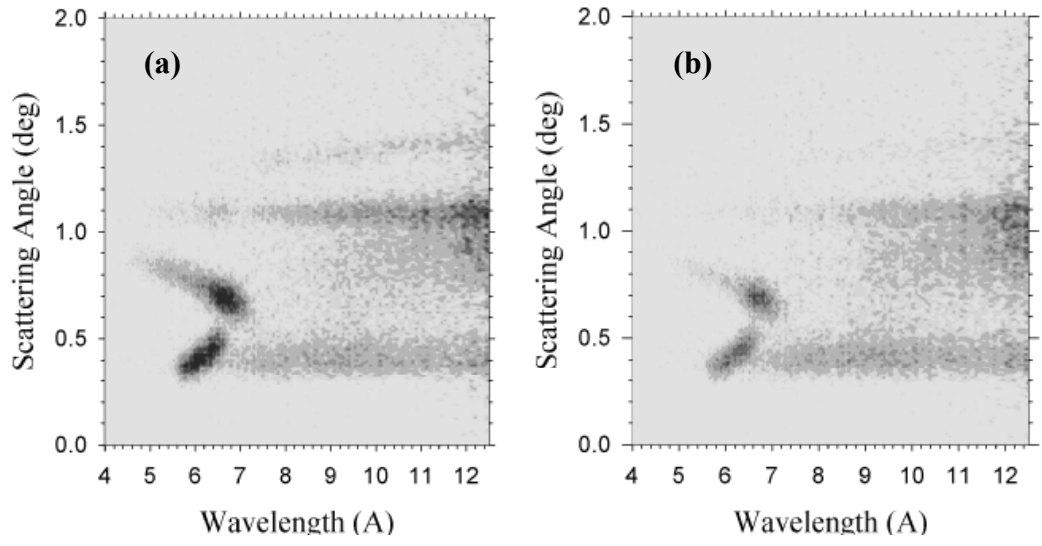

Fig. 1

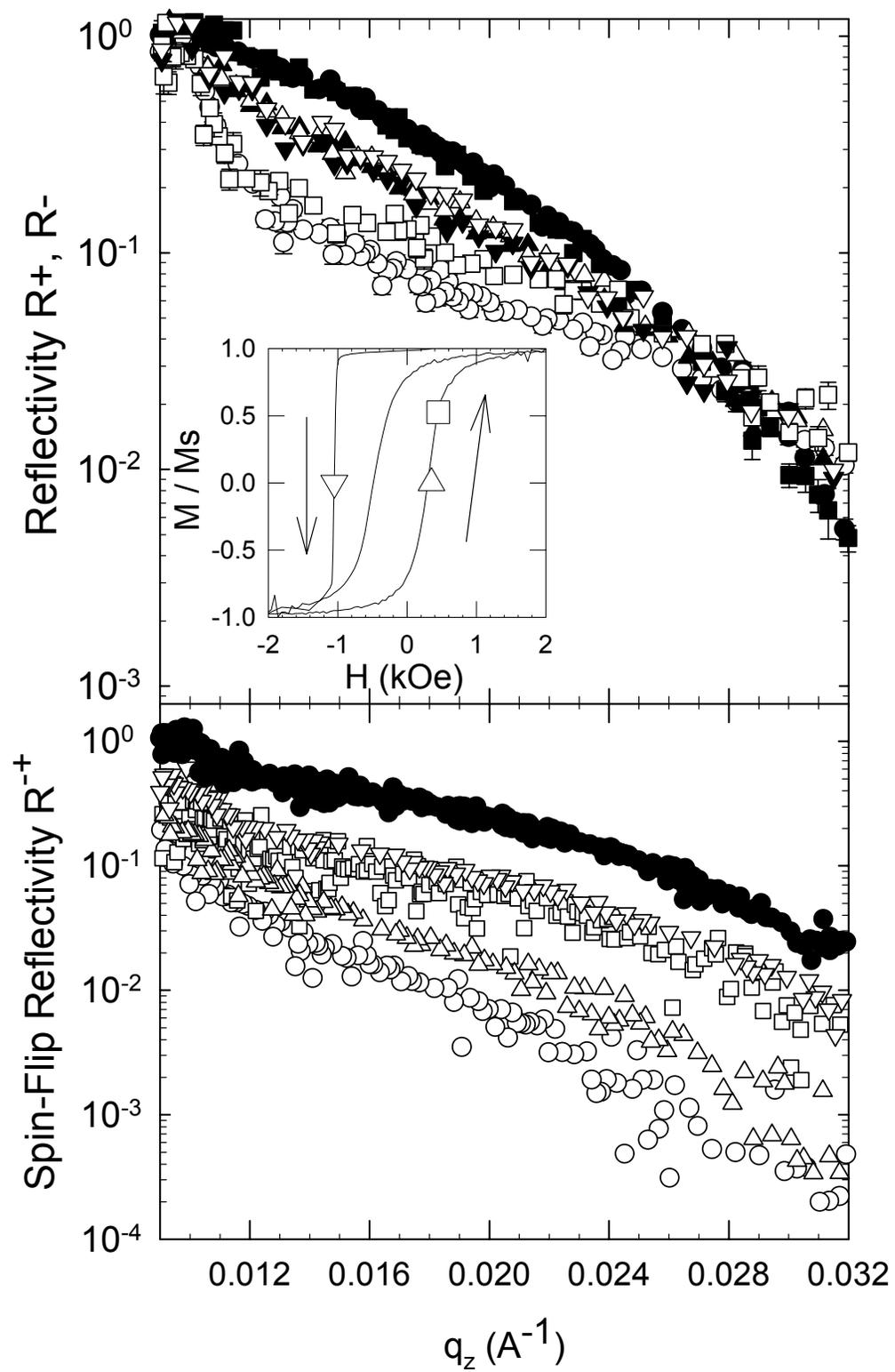

Fig. 2

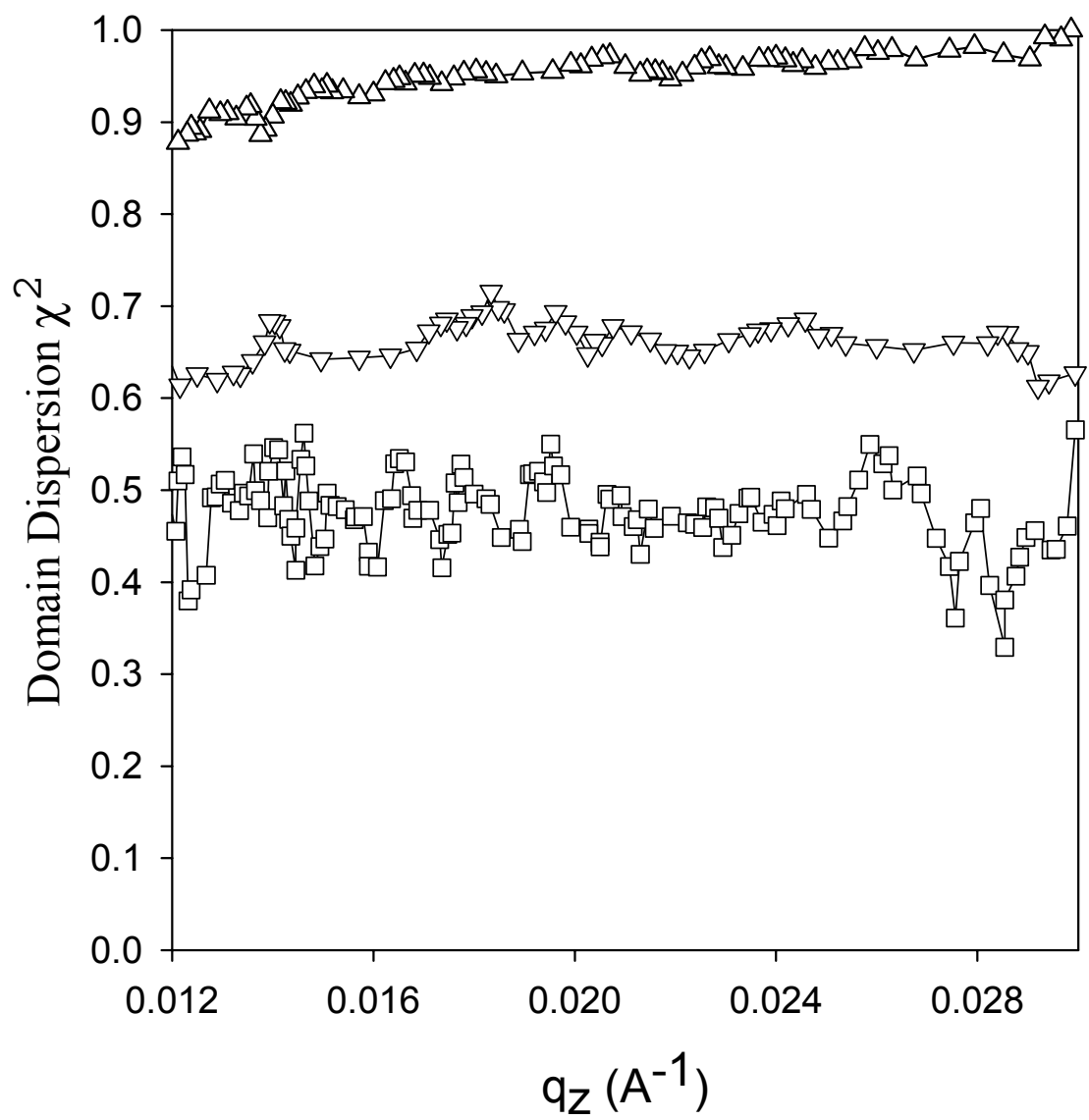